  \documentstyle{article}
\newcommand{\lab}{\label}
\newcommand{\bc}{\begin{center}}
\newcommand{\ec}{\end{center}}
\newcommand{\noi}{\noindent}
\newcommand{\barr}{\begin{array}}
\newcommand{\bey}{\begin{eqnarray}}
\newcommand{\be}{\begin{equation}}
\newcommand{\ear}{\end{array}}
\newcommand{\eey}{\end{eqnarray}}
\newcommand{\ee}{\end{equation}}

\newcommand{\pde}{\partial}

\newcommand{\spao}[1]{\mbox{\hspace{#1}}}
\newcommand{\spav}[1]{\parbox{1mm}{\vspace*{#1}}}

\newcommand{\ssty}{\scriptstyle}
\newcommand{\sssty}{\scriptscriptstyle}

\newsavebox{\ipiu}
\newsavebox{\imen}

\renewcommand{\a}{\alpha}
\renewcommand{\b}{\beta}

\newcommand{\da}{\dagger}

\newcommand{\D}{\Delta}

\renewcommand{\l}{\lambda}

\newcommand{\f}{\varphi}

\sbox{\ipiu}{$\ssty i \sssty +1$}
\sbox{\imen}{$\ssty i \sssty -1$}

\begin{document}

\begin{titlepage}

\rightline{{\bf DSF-T-94/17~~~~~}}

\rightline{{\bf INFN-NA-IV-94/17}}

\centering \spav{1cm}\\
{\LARGE\bf FQHE and Jain's approach on the torus\\}
\spav{2cm}\\
{\large G. Cristofano, G. Maiella, R. Musto and F. Nicodemi \\}
{\normalsize\em Dipartimento di Scienze Fisiche, Universit\`a di Napoli\\}
{\normalsize\em and INFN Sez. di Napoli\\}
\spav{3cm}\\
{\small\bf Abstract\\}
\spav{2mm}\\
{\small\parbox{13cm}{\spao{4mm}
By using
the explicit knowledge of the lowest energy single particle wave functions
 in the presence of an {\it arbitrary} magnetic field, we extend to the case
of a torus Jain's idea of looking at the FQHE as
a manifestation of an integer effect  for {\it composite
fermions}. We show that  this  can be realized
thanks to a redefinition of the vacuum state that is explicitly collective
in nature. We also discuss  the relationship of this approach
with the hierarchical scheme and with the characterization of the Hall
states  in terms of $W_{1+\infty}$ algebras and 2D conformal field theories.

}\\}
\vfill
{\large May 1994\\}
\spav{4mm}\\
Work supported in part by MURST and by EEC contract n. SC1-CT92-0789
\end{titlepage}

\large
\baselineskip=12pt

\bigskip
\bigskip

Since the fractional quantized Hall effect (FQHE)$^1$ was first observed by
Tsui, Stormer and Gossard in 1982 $^2$ considerable experimental progress has
been made, performing measurements with samples of higher mobility under
stronger magnetic fields and at lower temperature.

The experimental results relative to the sequence of filling factors
$\nu= p/(2p\pm 1)$ and  $\nu= p/(4p\pm 1)$ seem to give strong support to
the idea first suggested by Jain $^3$ of looking at the FQHE for electrons as
a manifestation of the integer effect (IQHE) for {\it composite
fermions}, obtained by attaching to each electron an even
number of flux units opposite to
the external magnetic field.  In fact,  not only the most prominent Hall
plateaux are seen at the fillings of the principal sequence $p/(2p\pm 1)$,
but also the energy gaps measured for this sequence correspond to the
cyclotron energies relative to the reduced magnetic field $B-B_{1/2}$ $^4$.
Furthermore, Halperin, Lee and Read $^5$, following a point of view closely
related to Jain's approach, have stressed the r\^ole of the state at $\nu
=1/2$,
arguing  for the existence of many features of a Fermi surface and computing
an anomaly in surface acoustic wave propagation in agreement with the results
of recent experiments $^6$.

The emerging picture for the FQHE seems to be quite simpler and clearer than
the traditional hierarchical scheme $^7$, which assigns the Hall plateaux
of the principal series to different levels of the hierarchy.  Still
two essential aspects of the FQHE, namely the presence of quasi-particle
vortex excitations, carrying fractional charge and statistics $^8$,
and the topological nature of
the Hall conductance $^9$, do not appear in a natural way in Jain's approach.
On
the other hand it is well known that both properties, at least for the
"Laughlin states" corresponding to fillings $\nu=1/m$,  are encoded in the
$m$-fold degeneracy of the electron wave function on the torus $^{10}$.
In fact the Hall conductance can be simply related to the behaviour of the
zeros
of the wave function under a change of boundary conditions and, due to
the $m$-fold degeneracy, one needs a change of phase between $0$ and $2m\pi$
along one the two cycles of the torus to achieve a full winding.
Furthermore, by using 2D conformal
field theory techniques, one can relate the $m$-fold degeneracy of the wave
function to the existence of $m$ distinct $g=1$ {\it
vacua}, each corresponding to a possible excitation with a charge multiple
of the elementary anyonic charge $1/m$ $^{11}$.

It is then a natural and relevant problem to analyze how the correct
multi-electron wave functions on the torus can be obtained within Jain's
approach, in view of the change of degeneracy going from integer to
fractional filling. In this paper we discuss this problem by using
the explicit knowledge of the lowest energy single particle wave functions
on the torus in the presence of an {\it arbitrary} magnetic field.
As a consequence of this simple and direct approach one can also
understand, already for the case of the plane, why it is possible
to look at the FQHE, that is a multi-particle
effect, as at an IQHE, usually thought of as a single particle phenomena,
for the composite fermions. This  can be realized
thanks to a redefinition of the vacuum state that is explicitly collective
in nature.

In the following we will be mainly concerned with the case $\nu
=1/m =1/(2k+1)$, corresponding to the filling of the first Landau level for
the composite fermions. We will briefly discuss the general case at the end
of the paper.

Let us start by recalling that for an arbitrary magnetic field orthogonal to
the $(x,y)$ plane in the gauge $\nabla \cdot \vec{A} =0$ one can
define  $\f(x,y)$ such that:

\be{e \over c\hbar} A_y = \pde _x \f \,\,\, \, \, \, ; ~~
{e \over c\hbar}A_x = -\pde _y \f
\lab{f}\ee \noi
and
\be{e \over c\hbar} B(x,y) = (\pde_x^2 + \pde _y ^2 ) \f \ee \noi

The Hamiltonian describing the motion of an electron in the
plane is given by the Pauli operator:

\be H=-{\hbar ^2\over 2m} \{ (\pde_x +i\pde_y \f )^2 +
(\pde_y -i\pde _x \f )^2 + {\sigma}_3 (\pde_x^2 +  \pde_y^2) \f\}
\ee  \noi
For an electron with spin aligned with the
magnetic field the states

\be \psi (x,y) = e^{-\f} \chi (z) \lab{l}\ee\noi
where $\chi(z)$ is an arbitrary analytic function of $z=x+iy$, have all zero
energy. These are actually the lowest energy states as the Hamiltonian is non
negative:
\be H={\hbar ^2\over m} a^{\da} a \ee \noi
where
\be a^{\da} = \sqrt{2} (-\pde + \pde \f)\,\,\, ; ~~~
a=\sqrt{2}( \bar{ \pde} +\bar{\pde} \f) \ee \noi
and, as usual, $\pde =\pde/\pde z \,\,\, ,~~\bar{\pde} =\pde/ \pde \bar{z}$.

If the magnetic field is uniform it is  convenient to work
in magnetic units $\l ^2 =c\hbar /(eB)$. Then $a$ and $a^{\da} $
are ordinary annihilation and creation operators: $[a,a^{\da} ]=1$.
For example in the symmetric gauge the potential is given by $\f_S
= {z\bar{z}  \over 4}$ and
\be a^{\da} = \sqrt{2} (-\pde +{\bar{z}\over 4 })\,\,\,\,\,\,
a=\sqrt{2} ( \bar{ \pde} +{z\over 4}) \ee \noi

\noindent It is also convenient to define the operators

\be b^{\da} =\sqrt{2}
 (-\bar{\pde} + \bar{\pde} \f_S) =
\sqrt{2} (-\bar{\pde} +{z\over 4}) \lab{b} \ee \noi
 \be b=\sqrt{2} (\pde +\pde \f_S) =\sqrt{2}( \pde +{\bar{z} \over 4}) \lab{b1}
\ee \noi
which act again as a couple of annihilation and creation
operators, $[b,b^{\da} ] =1$, commuting with $a$, with $a^{\da} $ and with the
Hamiltonian.
A complete set of states is then given by
\be \psi_{l,n} ={(a^{\da} )^l (b^{\da} )^n \over \sqrt{l!} \sqrt{n!}}
e^{-{z\bar{z} \over 4}}\ee

When dealing with the first Landau level one can consider $b$
and $b^{\da} $ as acting on the analytic part of the wave function, leading,
with suitable normalization, to
the Fock-Bargmann (FB) representation:
\be b= \pde \,\,\,\,\,\, \,\,\, b^{\da}  =z \lab{FB}\ee
Then, if $b\chi_0 =0$, a basis for the analytic part of the wave functions
can be written as:

\be \chi_n \,\, =\,\, {(b^{\da}  )^n \over \sqrt{n!}}\,\, \chi_0\,\,\, =
\,\, {z^n \over \sqrt{n!}} \lab{chi}
\ee \noi

For the Laughlin wave function at filling $\nu = 1$, describing a uniform
charge distribution in the thermodynamical limit, one has:

\be \chi_{\nu =1} (z_1, z_2, \dots z_{N_e})=\prod_{i<j} (z_i -z_j )
=\sum_P (-1)^P (b_1^{\da}  )^{n_{i_{1}}}
(b_2^{\da}  )^{n_{i_{2}}}  \dots (b^{\da}  _{N_e} )^{n_{i_{N_e}}}\chi_0
\lab{van}\ee \noi

Here $n_{i_1}, n_{i_2} \dots  n_{i_{N_e}}$ is
 a permutation, $P$, of the non negative integers smaller than
the electron number $N_e$.

Let us now study the motion of an electron in the presence of the uniform
magnetic field $B$ {\it and} of an infinitely thin magnetic flux tube
located at $z'$ of strength $-2k$ in units $\phi_0 = hc/e$.
The lowest energy states will still be of the
form given by eq. \ref{l} where the potential $\f$ can be taken as
\be \f= \f_S +h(z)= {z\bar{z}\over 4}-2k~ ln(z-z') \ee

The non-analytic part of the wave function is
unchanged with respect to the uniform case. Then, we can define $b'$ and
$b'^{\da}  $ along the same line of eqs. \ref{b} and \ref{b1}, namely:

\be b'^{\da}  = \sqrt{2}  (-\bar{\pde} + \bar{\pde} \f) =
\sqrt{2}  (-\bar{\pde} + \bar{\pde} \f_S )
\ee \noi
\be b'= \sqrt{2}( \pde +\pde \f)
=\sqrt{2}[ \pde +\pde (\f_S +h)] \lab{b'}
\ee \noi
They commute with $a$ and $a^{\da}$ up to delta-function terms and
their action on the analytic part is given by

\be b'= \pde +\pde h(z) = \pde -{2k \over z-z'}\,\,\, \,\,\, , ~~
b'^{\da}   =z \lab{FB'}\ee

\noindent The analytic part of the new states can then be written as
\be \chi '_n ={(b'^{\da}   )^n \over \sqrt{n!}}\chi '_0 \lab{chi'} \ee \noi
where the new vacuum state $\chi '_0 $ is defined by
 $b' \chi '_0 =0$, i.e. $\chi '_0 =(z-z')^{2k}$.

The presence of a zero of order $2k$ at $z'$ is expected, due to the
delta function singularity of the magnetic field. As a consequence, in
Jain's composite fermions approach, no zero is wasted $^{12}$ as flux tubes are
attached to the particles  and their presence plays the same r\^ole of a
singular repulsive two body potential. If at filling $\nu = 1/m= 1/(2k+1)$
the electrons are taken to be uniformly distributed with density $n_e$,
 the composite fermions see an effective magnetic field $B_{eff}= B-
2k\phi_0 n_e$ and have an effective filling $\nu_{eff}=1$.
More specifically, generalizing eq. \ref{FB'}, we define

\be b_i= \pde_i -2k\pde_i \sum_{j\neq i} ln(z_i-z_j) = \pde_i -2k\sum_{j\neq i}
{1 \over z_i-z_j}\,\,\, \,\,\, ,~~b_i^{\da}    =z_i \lab{FBi}\ee

Due to eq.
\ref{van},  \ref{chi'} and \ref{FBi}, the analytic part of the
wave function will be given by

\be \chi _{\nu ={1/m}} (z_1, z_2, \dots z_{N_e})=  \sum_P (-1)^P (b_1^{\da}
)^{n_{i_{1}}}
(b_2^{\da}   )^{n_{i_{2}}}  \dots (b_{N_e}^{\da}   )^{n_{i_{N_e}}}\chi_c =
\prod_{i<j} (z_i -z_j )^m \lab{VAN}\ee \noi
where $b_i \chi_c =0$ or

\be \chi_c= \prod_{i<j} (z_i -z_j )^{2k} \lab{chic} \ee \noi
Eq. \ref{VAN} gives the
Laughlin wave function at filling $\nu =1/m$ and the choice of the
{\it collective vacuum state} $\chi_c $ turns out to be consistent with the
requirement of uniform distribution.

Notice that for such a filling we could have directly bound $m$ units of
flux to the electrons turning them into {\it composite bosons} moving in an
average
zero field. We would have then defined the $b_i$ and $b^{\da}   _i$ operators,
as in
eq. \ref{FBi}, but with $2k$ turned into $m=2k+1$. The composite bosons would
condense in a collective vacuum state defined now by:
 \be \left ( \pde_i -m\sum_{j\neq i}
{1 \over z_i-z_j}\right) \chi (z_1, z_2, \cdots z_{N_e}) =0\lab{WZ}\ee\noi
Eq. \ref{WZ}, that leads again to the Laughlin wave function, can be seen
as the linear differential equation for the correlators of abelian Wess-Zumino
fields $^{13}$ of conformal weight $m/2$, making contact with the analysis
of the Hall effect in terms of 2D conformal field theory $^{14}$.

We close our discussion of Jain's approach on the plane
by recalling that the Laughlin state
at $\nu=1$ can be characterized as a highest weight vector of the
$W_{1+\infty}$ algebra of the area preserving non singular diffeomorfisms
 $^{15}$ generated by
\be {\cal L}_{m,n} = \sum _{i=1} ^{N_e} (b_i ^{\da} )^{m+1} (b_i )^{n+1}\,\,
\,,~  n,m \geq -1
\lab{calL} \ee \noi
where the $b_i$'s and $b_i^{\da} $'s are
in the FB representation, eq. \ref{FB}, that is
\be {\cal L}_{m,n} \chi_{\nu =1} (z_1, z_2, \dots z_{N_e}) =0 \, \, \, \, ,~
n>m\geq -1
\lab{calL0}  \ee

In the $\nu =1/m$ case  the operators $b_i$'s
and $b_i^{\da} $'s, given by eq. \ref{FBi}, satisfy creation and annihilation
commutation relations up to delta terms in the electrons relative coordinates:
\be [b_i ,b^{\da} _i]= 1+2k\pi \sum_{j\neq i} \delta (z_i -z_j) \, \, \, \, ,~~
[b_i ,b^{\da} _j]= -2k\pi  \delta (z_i -z_j) \,\,\, {j\neq i} \lab{bbdelta} \ee
However, acting on the collective vacuum they are algebraically equivalent
to those given by the FB representation \ref{FB}, as the contribution of the
delta terms in eq. \ref{bbdelta} will vanish. Taking into account the
structure of the Laughlin states, eqs. \ref{van} and \ref{VAN}, we
conclude that an algebraic characterization of incompressibility, analogous to
 eq. \ref{calL0}, still holds for filling $\nu =1/m$ $^{15,16}$, i.e.:
\be {\cal L}_{m,n} \chi_{\nu ={1 /m}} (z_1, z_2, \dots z_{N_e}) =0
\, \, \, \, \, \, n>m\geq -1 \lab{CALL0} \ee
where the ${\cal L}_{m,n} $ are still given by eq. \ref{calL} but in terms
of the new $b$'s.

2. In order to derive the Laughlin wave function on the torus $^{10}$ in the
framework of Jain's approach let us start by recalling the structure of the
doubly periodic  single particle wave functions in an uniform magnetic
field.

To make the argument as simple as possible we take a square torus of
length L in magnetic units, and, to have non trivial
solutions, we assume the total flux through the surface of the torus
to be an integer measured in quantum flux units, i.e. $ L^2=2 \pi
N_s$.  In the gauge $\vec{A}= y \hat{x}$ the eigenfunctions
of the first Landau level may be written in the form
\begin{equation}
\psi (x,y) = e^{-\frac{1}{2} y^2} f(z)
\end{equation}

As the vector potential has a discontinuity along the line
$y=L$, $\psi(x,y)$ is periodic under translation by $L$
in the $y$ direction only up to a gauge transformation:

\begin{equation}
\psi(x,y) \rightarrow \psi(x,y) e^{ixy}
\end{equation}
\noindent
Therefore we have the following periodicity conditions on $f(z)$:

\begin{equation}
f(z+L) = f(z) \,\,\,\,, ~~
f(z+ i L) = e^{\pi N_s} e^{-i2\pi{N_s\over L}z}\,\,f(z)  \lab{pc}
\end{equation}
Let us introduce the ``magnetic'' translation
operators $\cal{S}$ and $\cal{T}$, defined by:
\begin{equation}
{\cal S}_a f(z) = f(z+a)~~~~~~~~
{\cal T}_b f(z) = e^{-b^2/2 +ibz} f(z+ib)
\end{equation}
which satisfy the commutation relations of the Heisenberg group:

\begin{equation}
{\cal{S}}_a {\cal{T}}_b = e^{iab}{\cal{T}}_b {\cal S}_a
\end{equation}

The independent solutions of the periodicity conditions, eq.\ref{pc},
 form a basis for
the $N_s$-dimensional representation of the discrete subgroup of magnetic
translations generated by ${\cal{S}}_{L/N_s}$ and  ${\cal{T}}_{L/N_s}$  and
can be taken as the Theta-functions with rational characteristics $l/N_s$
where $l= 1,2, ..., N_s$ $^{17}$:
\bigskip
\begin{equation}
f_l (z) = \Theta \Bigg[ \begin{array}{c} l/N_s \\[-2mm] 0 \end{array} \Bigg]
\left (z\frac{N_s}{L}|iN_s \right ) \lab{f_l}
\end{equation}

This discussion can be easily generalized $^{18}$ to the case of an arbitrary
magnetic field, provided the flux quantization condition is verified. We will
not enter into an analysis of this point that will become evident from the
discussion of the specific case of interest given below.

It is also  interesting to establish  the connection of the states
given by eq. \ref{f_l} with
a set of {\it generalized coherent states} $^{19}$. Define the generators of
the
generalized coherent states as

\begin{equation}
{\cal D}_l (b,b^{\da} |N_s) = \sum_{r \in Z} e^{(\a_r^{l}b^{\da}
-\bar{\a}_r^{l}b)}=
\sum_{r \in Z} e^{-{|\a_r^l|^2 \over 2}}e^{\a_r^{l}b^{\da} }e^{
-\bar{\a}_r^{l}b}
\lab{coh} \ee \noi
where $\a_r^{l} =iL(r+l/N_s)$ and $b$, $b^{\da} $ are in the FB representation,
eq \ref{FB}. Then

\be f_l (z) = {\cal D}_l(b,b^{\da} |N_s) \chi_0 \lab{coh0}\ee \noi
where $b \chi_0=0$.

Then, at filling $\nu =N_e/N_s=1$,
the analytic part of the Laughlin wave function on the torus
 can be written as

\be
 f_{\nu=1} (z_1 , z_2, \dots , z_{N_e})= \sum_P (-1)^P
{\cal D}_{i_1}(b_1,b^{\da}_1 |N_e) {\cal D}_{i_2}(b_{2},b^{\da} _{2}|N_e)
\dots {\cal D}_{i_{N_e}}(b_{N_e},b^{\da} _{N_e}|N_e) \chi_0 \lab{det} \ee

Here $i_1, i_2 \dots  i_{N_e}$ is  a permutation, $P$, of
the sequence of natural numbers not bigger than the electron number $N_e$.

On the basis of the periodicity properties of the $\Theta $'s and of the
location of the zeros it can be seen that\footnote{For simplicity we limit
ourselves to $N_e$ odd. The case $N_e$ even would require only a
slight generalization of eq. \ref{nu1}.}
\be
 f_{\nu=1} (z_1 , z_2, ...z_{N_e})=
\Theta \Big[  \begin{array}{c}1\\[-2mm]0 \end{array} \Big]
\left (\frac{Z}{L}|i \right ) \prod_{i<j} \Theta_1 (z_{ij} | i)
\lab{nu1} \ee
\noindent
where $Z= \sum_{i=1}^{N_e}  z_i $ is the "center
of charge" coordinate, $z_{ij} = (z_i -z_j)/L $ and $\Theta_1 =$ ~~~
$\Theta {\small \Big[  \begin{array}{c}1/2\\[-2mm]1/2 \end{array} \Big] }$.

Let us now add to the previous uniform magnetic field an infinitely thin
flux tube located at $z'$ of strength $-2k$. Then the single particle wave
functions may be written as

\begin{equation}
\psi (x,y) = e^{-\frac{1}{2} y^2+ h(z)} g(z)
\end{equation} \noi
where, with suitable normalization,
\be h(z) =2k~ln
\frac{\Theta_1 ({z-z'\over L} | i)}{{\Theta '}_1 (0 | i )} \lab{h(z)}\ee
Taking into account the transformation properties
\be \Theta \Big[  \begin{array}{c}\a\\[-2mm]\b \end{array} \Big]
\left (w +1|\tau  \right ) =e^{2\pi i \a}
\Theta \Big[  \begin{array}{c}\a\\[-2mm]\b \end{array} \Big]
\left (w |\tau  \right )
\lab{thtr}\ee

\be \Theta \Big[  \begin{array}{c}\a\\[-2mm]\b \end{array} \Big]
\left (w +\tau|\tau  \right ) =e^{-2\pi i \b}e^{-\pi i \tau}e^{-2\pi i w}
\Theta \Big[  \begin{array}{c}\a\\[-2mm]\b \end{array} \Big]
\left (w |\tau  \right )
\lab{ithtr}\ee \noi

\noindent one gets the following boundary conditions for  $g$

\begin{equation}
g(z+L) = g(z)
\end{equation}

\begin{equation}
g(z+ i L) = g(z)  e^{\pi (N_s-2k)} e^{-i2\pi{N_s\over L}z}
e^{i2\pi{2k\over L}(z-z')}
\end{equation}
which, due to eqs. \ref{thtr} and \ref{ithtr}, are satisfied by

\begin{equation}
g_l (z) = \Theta \Big[ \begin{array}{c} l/(N_s-2k)\\[-2mm]0 \end{array}
\Big]
\left (z\frac{N_s-2k}{L}+z'\frac{2k}{L}|i(N_s-2k) \right )
\end{equation}

When we extend the discussion to the system of Jain's composite fermions and
study the motion of the $i^{th}$ particle in the presence of all
the others we have

\begin{equation}
\psi (x_i,y_i) = e^{-\frac{1}{2} y_i^2+ h(z_i)} g(z_i) \lab{ps_i}
\end{equation} \noi
where
\be h(z_i) =2k \sum_{j\neq i} ln \left [
\frac{\Theta_1 (z_{ij} | i)}{{\Theta '}_1 (0 | i )} \right ]\lab{h(zi)}\ee
with the following boundary conditions

\begin{equation}
g(z_i+L) = g (z_i)
\end{equation}

\begin{equation}
g (z_i + i L) = g (z_i)  e^{\pi N_e}  e^{-i2\pi{N_e\over L}z_i}
e^{\pi 2k} e^{-i2\pi{2k\over L} Z} \lab{gzi}
\end{equation}

It is clear from eqs. \ref{ps_i} and \ref{gzi} that the wave function of
each composite fermion explicitly depend on the position of all the others.
To obtain an expression like eq. \ref{det} for the
composite fermions, but corresponding to a filling $\nu_{eff} =1$  as expected
in Jain's picture, we introduce a collective vacuum
\be \chi_c(z_1, z_2, \cdots z_{N_e})=\chi(Z) \prod_{i\neq j}  \left [
\frac{\Theta_1 (z_{ij} | i)}{{\Theta '}_1 (0 | i )} \right ]^{2k}
\lab {Chic}\ee
or, in other terms, we introduce the operators $b_i$ and $b_i^{\da} $
\be b_i= \pde_i -2k \pde_i \sum_{j\neq i} ln\left [
\frac{\Theta_1 (z_{ij} | i)}{{\Theta '}_1 (0 | i )} \right ]-\pde_i ln
\chi(Z)\,\,\,\,, ~~b_i^{\da}  =z_i \lab{FBit}\ee
so that $b_i \chi_c=0$.  The dependence on the center of charge variable,
that we have left unspecified,
must be consistent with eq. \ref{gzi} and with the requirement that
the multi-particle wave
functions lead to a uniform charge distribution. These wave functions take
now the same form as for ordinary fermions at $\nu =1$, eq. \ref{det}, i.e.

\be f (z_1 , z_2 ...z_{N_e})= \sum_P (-1)^P
{\cal D}_{i_1}(b_{1},b^{\da} _{1} |N_e) {\cal D}_{i_2}(b_{2},b^{\da} _{2}|N_e)
\cdots {\cal D}_{i_{N_e}}(b_{N_e},b^{\da} _{N_e}|N_e) \chi_c \lab{DET} \ee \noi
where the $D$'s are still given by eq. \ref{coh} with
$b_i$ and $b^{\da} _i$ given by eq. \ref{FBit} and $\chi_c$ by eq.
\ref{Chic}. Taking into account eq. \ref{nu1} we have

\be
f(z_1 , z_2 ...z_{N_e})= \prod_{i<j=1}^{N_e} \left [
\frac{\Theta_1 (z_{ij} | i)}{{\Theta '}_1 (0 | i )}
\right ]^m F(Z)  \lab{nu1m} \ee \noi
where
\be F(Z) = \chi(Z)
\Theta \Big[  \begin{array}{c}1\\[-2mm]0 \end{array} \Big]
\left (\frac{Z}{L}|i \right ) \ee
\noindent
The analytic part of the multi-particle wave functions, eq. \ref{nu1m},
 will give the prescribed periodicity condition, eq. \ref{pc}, in each
variable provided that:

\begin{equation}
F(Z+L) = F(Z) \,\,\,\,,~~
F(Z+ i L) = F(Z) e^{\pi m} e^{-i2\pi{m\over L}Z} \lab{PC}
\end{equation} \noi

On the other hand these conditions on $F(Z)$ imply a uniform charge
distribution. Indeed one easily sees that eqs.
\ref{PC} require that the charge density is invariant, both in the $x$ and
$y$ directions, under a translation of $\D =L/N_e$, which is vanishing
small in the thermodynamical limit.

Then the $\nu_{eff}=1$ wave function for Jain's composite fermions leads  to
the usual multi-electron wave functions on the torus at filling $\nu=1/m$
$^{10}$,
as eqs.\ref{PC} are satisfied by the $m$ independent functions:

\be F_l (Z)= \Theta \Big[  \begin{array}{c}l/m\\[-2mm]0 \end{array} \Big]
\left (\frac{Zm}{L}|i m \right ) \lab{ccm} \ee

The novel feature of the degeneracy arises from the
different possible choices of the dependence on the center of charge variable
consistent with the requirement of an uniform charge distribution.

3. We close this paper with a brief discussion of the FQHE at arbitrary
rational filling, limiting ourselves to the case of the plane. Our simple
 analysis  of the  electron motion in the presence
of a uniform magnetic background and localized flux tubes can be used
to discuss the FQHE at fillings of the form $\nu = p/(2kp+1)$ as IQHE for
Jain's composite fermions at integer filling $p$. Indeed the
appropriate operators $a$, $a^{\da}$  and $b$, $b^{\da}$ have commutation
relations that differ from those relative to a purely uniform
background only by terms that have delta functions in the relative
coordinates. As they will be acting on the appropriate collective vacuum
states, such as \ref{chic}, the Landau level structure will be preserved and
the wave function for composite fermions at filling $p$ will differ by the
ordinary one only by a factor that has a zero of order $2k$ in each
relative coordinate. As shown by Jain such a state lies predominantly in the
lowest Landau level for $N_e$ large $^2$.

On the other hand this approach can be extended in a way that mimics the
traditional hierarchical scheme, by introducing a number of charged
excitations $N_I$ at the $I^{th}$ hierarchical level, each carrying its own
statistical flux tube. One is then led to introduce a matrix $K_{IJ}$,
describing the couplings  of a particle at $I^{th}$ level with the
statistical field relative to the  $J^{th}$ level or, in other terms, the
braiding factors between the corresponding "particles". The discussion of
the motion of the $i^{th}$ "particle" of the $I^{th}$  level in the
presence of an external magnetic field and of flux tubes located at the
positions of all the other "particles" leads naturally to
the introduction of the operators:

\be (b_i^I)^{\da} =z_i^I\,\,\, \,\,\, ,~
b_i^I= \pde_i^I -\pde_i^I\,\, {\sum}^{'} H_{IJ}\,\, ln(z_i^I-z_j^J) =\pde_i^I
 -\sum_{(j,J)\neq (i,I)}{H_{IJ}\over z_i^I-z_j^J}  \lab{FBI}\ee \noi
where $H_{IJ} = K_{IJ}-${\bf 1}, $i$ and $j$ label the "particles" in a
given level while $I$ and $J$ run on the different levels of the hierarchy
and $\sum^{'}$ means the sum on all possible couples.

Then one can introduce a "collective vacuum"  $\chi_H$ defined by
$b_i^I \,\chi_H=0$ and write a "wave function" along the same line of eq.
\ref{van} and \ref{VAN}, namely

\be \chi  (\{z_i^I\})= \prod_I \sum_{P_I}(-1)^{P_I} (b_1^{I\da})^{n^I_{i_1}}
(b_2^{I \da})^{n^I_{i_2}}  \dots (b_N^{I \da})^{n^I_{i_{N^I}}}\chi_H
\lab{VANH}\ee \noi
where $n^I_{i_1}, n^I_{i_2} \dots
n^I_{i_{N^I}}$ is
 a permutation, $P^I$, of the sequence of non negative integers smaller than
the
number $N^I$ of "particles" at level $I$.

It can be seen that this simple approach is in agreement with the general
description of the hierarchical model given by means of an abelian
Chern-Simons field theory $^{20}$ or by using 2D Coulomb gas Vertex operators
$^{21}$.
Furthermore, it allows to verify that all the hierarchical states can be
classified according to the universality class of a $W_{1+\infty}$ algebra
$^{22}$,generalizing the definition of the generators, eq. \ref{calL}, $^{16}$:

\be {\cal L}_{m,n} = \sum _{i,I} (b_i ^{I\da} )^{m+1} (b_i^I )^{n+1}\,\,\, ,~~
 n,m \geq -1
\lab{CALLL} \ee \noi

All these different approaches, describing  the FQHE in terms of an
appropriate set of quasiparticles and relative statistical fields,
whose properties are encoded
in the matrix of their braiding factors,  are essentially kinematical in
nature, as they give all possible quantum Hall fluids. From this point of
view the reason why the sequence of fillings selected by Jain's approach
appear experimentally as the most prominent one is left unanswered and
should be addressed at a dynamical level.

{\bf Acknowledgements}

Discussions with Y.S. Wu on different aspects of FQHE are gratefully
acknowledged.

\begin{center}
{\bf References}
\end{center}

\smallskip

\begin{enumerate}

\item  A useful collection of reprints relative to the FQHE with
interesting introduction to the different theoretical approaches can be
found in the volume {\it Quantum Hall Effect} Ed. Michael Stone,
World Scientific Singapore 1992. We denote by a
$~^*$ all the papers reprinted therein.

\item D.C. Tsui, H. L. Stormer and A. C. Gossard, Phys. Rev. Lett. {\bf
48}, 1559 (1982).

\item J. K. Jain, Phys. Rev. Lett. {\bf 65}, 199 (1989)$~^*$ ; Phys. Rev. {\bf
B 40} 8079 (1989);{\bf B 41} 7653 (1990).

\item  R.R. Du, H. L. Stormer, D.C. Tsui, L. N. Pfeiffer  and  K. W. West,
Phys. Rev. Lett. {\bf 70}, 2944 (1993).

\item B. I. Halperin, P. A. Lee and N. Read, Phys. Rev. {\bf B 47} 7312
(1993).

\item  R. L. Willett, M.A. Paalanen, R. R. Ruel,  K. W. West, L. N. Pfeiffer
and D. J.
Bishop Phys. Rev. Lett. {\bf 65}, 112 (1990);
 R. L. Willett, R. R. Ruel, M. A. Paalanen,  K. W. West and
L. N. Pfeiffer  Phys. Rev. {\bf B  47}, 7344  (1993).

\item F. D. M. Haldane, {\it Phys. Rev. Lett.} {\bf  51} 605 (1983)$~^*$;
B. I. Halperin, {\it Phys. Rev. Lett.} {\bf 52} 1583; 2390(E)(1984)$~^*$.

\item R. B.  Laughlin, Phys. Rev. Lett. {\bf 50} 1395 (1983)$~^*$;
D. Arovas, J. R. Schrieffer, F. Wilczek, Phys. Rev. Lett. {\bf 53}
722 (1984)$~^*$; S. M. Girvin articles contained in {\it The Quantum Hall
Effect}, ed. R. E. Prange and S. M. Girvin, Springer, New York 1987.

\item For a general introduction to the topological aspects of the QHE see,
e.g., Chap. 3 of ref. 1 and the reprints and references contained therein.

\item F.D.M. Haldane and E.H. Rezayi, {\it Phys. Rev.} {\bf B31} 2529
(1985)$~*$; R. B. Laughlin, {\it Ann. of Phys.} {\bf 191} 163 (1989).

\item G. Cristofano, G. Maiella, R. Musto and F. Nicodemi, {\it Phys. Lett.}
{\bf 237B} 379 (1990).

\item B. I. Halperin, Helv. Phys. Acta {\bf 56} 75(1983).

\item V. G. Knizhnik and A. B. Zamolodchikov, Nucl. Phys. {\bf B247} 83 (1984).

\item  For a review see G. Cristofano, G. Maiella, R. Musto and F. Nicodemi,
Nucl. Phys. {\bf 33C} 119 (1993) and references contained therein.

\item A. Cappelli, C. A. Trugenberger and G. R. Zemba, Nucl. Phys. {\bf B396}
465 (1993); Phys. Lett. {\bf 306 B} 100 (1993).

\item M. Flohr and R. Varnhagen, preprint BONN-HE-9329; hep-th/9309083 1993.

\item D. Mumford, {\it Tata Lectures on Theta} {\bf I}, Birkh\"{a}user Ed.,
Boston 1983.

\item B. A. Dubrovin and S. P. Novikov, Sov. Phys. JETP {\bf 53} 511 (1980).

\item A. Perelomov, {\it Generalized Coherent States and their
Applications}, Springer-Verlag, Berlin, 1986.

\item J. Fr\"olich and A. Zee, Nucl. Phys. {\bf 354 B} 1991 369;
X.-G. Wen and A. Zee, Phys. Rew.  {\bf 46 B} 2290 (1993).

\item Cristofano, G. Maiella, R. Musto and F. Nicodemi, Modern Phys. Lett.
{\bf 7 A} 2583 (1992).

\item A. Cappelli, C. A. Trugenberger and G. R. Zemba, Phys. Rev. Lett. {\bf
72} 1902 (1994).

\end{enumerate}

\end{document}